\documentclass[aps,superscriptaddress,eqsecnum,nofootinbib,showpacs,preprintnumbers]
{revtex4}
\usepackage{graphicx,epsfig}
\usepackage{amsmath}
\usepackage {amssymb}

\newcommand{\be}{\begin{eqnarray}}
\newcommand{\ee}{\end{eqnarray}}
\newcommand{\bea}{\begin{eqnarray}}

\newcommand{\eea}{\end{eqnarray}}

\makeindex

\begin{document}

\title{AdS and Lifshitz black hole solutions in conformal gravity sourced with a scalar field}
\author{Felipe Herrera}
\email{fherrera@userena.cl}
\affiliation{Departamento de F\'isica y Astronom\'ia, Facultad de Ciencias, Universidad de La Serena,\\
Avenida Cisternas 1200, La Serena, Chile.}

\author{Yerko V\'asquez}
\email{yvasquez@userena.cl}
\affiliation{Departamento de F\'isica y Astronom\'ia, Facultad de Ciencias, Universidad de La Serena,\\
Avenida Cisternas 1200, La Serena, Chile.}
\date{\today }

\begin{abstract}

In this paper we obtain exact asymptotically anti-de Sitter black hole solutions and asymptotically Lifshitz black hole solutions with dynamical exponents $z=0$ and $z=4$ of four-dimensional conformal gravity coupled with a self-interacting conformally invariant scalar field. 
Then, we compute their thermodynamical quantities, such as the mass, the Wald entropy and the Hawking temperature. The mass expression is obtained by using the generalized off-shell Noether potential formulation. It is found that the anti-de Sitter black holes as well as the Lifshitz black holes with $z=0$ have zero mass and zero entropy, although they have non-zero temperature. A similar behavior has been observed in previous works, where the integration constant is not associated with a conserved charge, and it can be interpreted as a kind of gravitational hair. On the other hand, the Lifshitz black holes with dynamical exponent $z=4$ have non-zero conserved charges, and the first law of black hole thermodynamics holds. Also, we analyze the horizon thermodynamics for the Lifshitz black holes with  $z=4$, and we show that the first law of black hole thermodynamics arises from the field equations evaluated on the horizon. Furthermore, we study the propagation of a conformally coupled scalar field on these backgrounds and we find the quasinormal modes analytically in several cases. We find that for anti-de Sitter black holes and Lifshitz black holes with $z=4$, there is a continuous spectrum of frequencies for Dirichlet boundary condition; however, we show that discrete sets of well defined quasinormal frequencies can be obtained by considering Neumann boundary conditions.

\end{abstract}

\maketitle

\newpage
\tableofcontents

\newpage
\section{Introduction}

An interesting class of space-times that have received considerable attention from the point of view of condensed matter physics are the Lifshitz space-times, which are described by the metrics \cite{Kachru:2008yh}
\begin{equation}
ds^2=- \frac{r^{2z}}{\ell^{2z}}dt^2+\frac{\ell^2}{r^2}dr^2+\frac{r^2}{\ell^2} d\vec{x}%
^2~,  \label{lif1}
\end{equation}
where $\vec{x}$ represents a $D-2$ dimensional spatial vector, with $D$ corresponding to the
space-time dimension and $\ell$ denoting the length scale in this geometry. These geometries are characterized by the parameter $z$ known as the dynamical or critical exponent. For $z=1$ this space-time is the usual anti-de Sitter (AdS) metric in Poincar\'e coordinates with the AdS radius $\ell$. According to the AdS/CFT correspondence \cite{Maldacena:1997re} and its generalizations beyond high energy physics, these metrics are important in the search for gravity duals of systems that appear in condensed matter physics and in quantum chromodynamics \cite{Kachru:2008yh, Hartnoll:2009ns}. There are many theories of
interest when studying such critical points, such theories exhibit the
anisotropic scale invariance $t\rightarrow \lambda ^{z}t$, $\vec{x}  \rightarrow \lambda \vec{x}$, where the dynamical exponent $z$ accounts for the different scale transformation between the temporal and spatial coordinates,
and they are of particular interest in studies of critical exponent theory
and phase transitions. The Lifshitz space-time exhibits these symmetries along with the scaling of the radial coordinate $r \rightarrow \lambda^{-1} r$. On the other hand, gravity duals of field theories with an anisotropic scale invariance at a finite temperature are represented by black hole geometries whose asymptotic behavior is given by (\ref{lif1}), known as asymptotically Lifshitz black holes. Several Lifshitz black hole metrics have been found, which are solutions of Einstein gravity with several matter fields and also of gravity theories in vacuum with higher curvature corrections, such as $f(R)$, Lovelock, New Massive Gravity and others (see for instance \cite{AyonBeato:2009nh, Cai:2009ac, AyonBeato:2010tm, Dehghani:2010kd, Mann:2009yx, Balasubramanian:2009rx, Bertoldi:2009vn, Bravo-Gaete:2013dca, Correa:2014ika, Alvarez:2014pra}).

 Asymptotically Lifshitz black holes can also emerge in Conformal Weyl Gravity in four dimensions with dynamical exponents $z=0$ and $z=4$ \cite{Lu:2012xu}. Conformal Gravity (CG) in four dimensions is a higher derivative theory of gravity, whose action is given by the square of the Weyl tensor, and it is invariant under conformal transformations of the metric tensor:
\begin{equation}\label{conform}
g_{\mu \nu}(x) \rightarrow \Omega^2(x)  g_{\mu \nu} (x)~,
\end{equation}
where the conformal factor $\Omega$ is a function of the coordinates. A consequence of this is that CG is sensitive to angles but not to distances. CG is naturally subject to conformal anomaly, and renormalizability requires the inclusion of Ricci squared terms. On the other hand, Einstein gravity in four dimensions is two-loop divergent \cite{Goroff:1985th}, and is perturbatively renormalizable when extended by the inclusion of curvature squared terms in the Lagrangian \cite{Stelle:1976gc, Stelle:1977ry}
; moreover, in contrast to Einstein gravity, CG contains ghostlike degrees of freedom in the form of massive spin-2 excitations, which implies vacuum instability. However, Einstein gravity with a cosmological constant is equivalent, at the tree level, to four-dimensional CG with Neumann boundary conditions \cite{Maldacena:2011mk}. The equivalence of the two theories, CG and Einstein gravity,  was also studied in \cite{Anastasiou:2016jix}, both at the level of the action and at the variation of it. 
Furthermore, an important characteristic is that any space-time conformal to an Einstein manifold is a solution of CG and, because the equations of motion of CG contain fourth order derivatives, there are other solutions to CG than just Einstein gravity. Another connection between both theories is that the renormalized action of four-dimensional Einstein gravity in asymptotically hyperbolic Einstein spaces is on-shell equivalent to the action of CG \cite{and}. 
The most general spherically symmetric asymptotically AdS solution of four-dimensional CG minimally coupled with an electromagnetic field is static and was obtained in \cite{Riegert:1984zz}. Moreover, it has been argued that CG may be able to explain the galactic rotation curves without the assumption of dark matter \cite{Mannheim:2005bfa}. Further black hole solutions to the field equations of four-dimensional CG in vacuum and in the presence of a Maxwell field include the general Kerr-NUT-AdS-like solution \cite{Mannheim:1990ya}, charged and rotating AdS black holes \cite{Liu:2012xn}, cylindrical black hole solutions \cite{Said:2012pm}, and also black hole solutions with both Abelian electromagnetic and $SU(2)$ Yang-Mills fields were obtained in \cite{Fan:2014ixa}. Another consequence of the invariance of CG under the transformation (\ref{conform}) is that the gravitational field equations of CG, which in four space-time dimensions are given by the vanishing of the Bach tensor, imply that the trace of the energy-momentum tensor of the matter sources must vanish. This is because the Bach tensor is traceless; therefore, among the matter sources which can be coupled to CG, one must consider conformally invariant fields such as a minimally coupled electromagnetic field, SU(2) Yang-Mills fields, or a conformally invariant scalar field. It is worth stressing that only a few solutions to CG with a conformally coupled scalar field have been obtained in the literature. Among them, we can mention boson stars with a self-interacting complex scalar field \cite{Brihaye:2009ef} and a C-metric with a conformally coupled scalar field \cite{Meng:2016gyt}. On the other hand, conformally invariant theories of gravity also exist in other space-time dimensions. In three space-time dimensions the equations of motion of CG contain third derivatives of the metric, and some black hole solutions have been studied in \cite{Oliva:2009hz, Guralnik:2003we, Grumiller:2003ad}. Additionally, three-dimensional Lifshitz black holes with $z=0$ were obtained in \cite{Catalan:2014una}. In six space-time dimensions, AdS and Lifshitz black hole solutions of CG in vacuum and also coupled to conformal matter were obtained in \cite{Lu:2013hx}.

 According to the AdS/CFT correspondence, the search for black hole solutions to theories of gravity coupled with scalar fields and electromagnetic fields is an important task in the studies of holographic superconductors \cite{Hartnoll:2008vx}, which is dual to a system consisting of a black hole with a scalar field minimally coupled to gravity, which below a critical temperature condenses outside the black hole horizon. The dual description of a superconductor also works for scalar fields non-minimally coupled to gravity \cite{Kuang:2016edj}. Further, Lifshitz holographic superconductivity has been the topic of numerous studies with interesting properties when the duality principle is generalized to non-relativistic systems \cite{Hartnoll:2009ns, Brynjolfsson:2009ct}. In addition to the recent applications of the gauge/gravity duality to condense matter phenomena like superconductivity, the behavior of scalar fields outside black holes has been extensively studied over the years, mainly in connection with the no-hair theorems (for a review on this topic see \cite{Sotiriou:2015pka, Herdeiro:2015waa, Volkov:2016ehx}). The basic physical requirement of a black hole with scalar hair is the scalar field must be regular on the horizon and fall off sufficiently rapidly at infinity.  From the above discussion, we conclude that it is important to understand the behavior of matter fields all the way from the black hole horizon to asymptotic infinity.

An important characteristic of black holes is their proper (quasinormal) oscillations. The quasinormal modes (QNMs) have a long history \cite{Regge:1957td, Zerilli:1971wd, Zerilli:1970se, Kokkotas:1999bd, Nollert:1999ji, Konoplya:2011qq}, and nowadays are of great interest due to the observation of gravitational waves from the merger of two black holes \cite{Abbott:2016blz}. The QNMs give information about the stability of the black holes and also of probe matter fields that evolve perturbatively in the exterior region of black holes, without backreaction on the metric. Also, the QNMs are important in the context of the gravity/gauge duality, because they give information about the relaxation times of thermal states in the boundary theory \cite{Horowitz:1999jd}, where the relaxation time is proportional to the inverse of the smallest imaginary part of the quasinormal frequencies (QNFs). On the other hand, in the context of black hole thermodynamics, the quantum area spectrum \cite{Bekenstein:1974jk} as well as the mass and entropy spectrum can be determined through the knowledge of the QNMs \cite{Hod:1998vk, Kunstatter:2002pj,Maggiore:2007nq}. The QNMs for asymptotically Lifshitz black holes have been studied for different types of perturbations. The QNMs for scalar field perturbations on the background of Lifshiftz black holes have been analyzed in \citep{CuadrosMelgar:2011up, Gonzalez:2012de, Gonzalez:2012xc, Myung:2012cb, Becar:2012bj,Giacomini:2012hg, Lepe:2012zf, Catalan:2014ama, Catalan:2014una, Sybesma:2015oha, Becar:2015kpa, Zangeneh:2017rhc}, and generally the scalar modes are stable. Also, the propagation of fermionic perturbations and the QNMs of Lifshitz black holes were studied in Refs. \cite{CuadrosMelgar:2011up, Gonzalez:2017ptj, Catalan:2013eza, Lopez-Ortega:2014daa}.  The electromagnetic QNMs were investigated in \cite{Lopez-Ortega:2014oha}. Furthermore, the QNMs of Lifshitz black holes with hyperscaling violation have been considered in Refs. \cite{BAI:2013koa, Gonzalez:2015gla, Becar:2015gca}.

The aim of this paper is to find asymptotically AdS black hole solutions and asymptotically Lifshitz black hole solutions to a gravitational system, which for simplicity consists only of a real scalar field conformally coupled to CG in four dimensions, which is an important issue due to their potential applications in the description of dual field theories at a finite temperature through the gauge/gravity duality to condense matter phenomena. We require the scalar field to be regular on the horizon and to fall off at spatial infinity. We also assume spherical symmetry of the system. Then, the thermodynamical properties of the solutions are studied, such as their mass, entropy and temperature and it is verified if the first law holds. Furthermore, we study the propagation of a conformally coupled test scalar field in the black hole solutions found and determine the QNMs analytically in several cases. Remarkably, such as in the free source case, we find that there are analytical AdS black hole and Lifshitz black hole solutions with dynamical exponents $z=0$ and $z=4$. On the other hand, calculating conserved charges of Lifshitz black holes is a difficult task; however, progress was made on the computation of the mass and the related thermodynamics quantities by using the Abbott-Deser-Tekin (ADT) method \cite{Devecioglu:2010sf, Devecioglu:2011yi}  and its off-shell generalization \cite{Gim:2014nba, Ayon-Beato:2015jga}, as well as the Euclidean action approach \cite{Gonzalez:2011nz, Myung:2012cb}. In order to calculate the conserved charges, we employ the generalized ADT formalism proposed in \cite{Kim:2013zha}, which corresponds to the off-shell generalization of the on-shell Noether potential of the ADT formalism \cite{Abbott:1981ff, Deser:2002rt}. We derive an expression for the contribution of the scalar field to the Noether potential, and we show that the method gives consistent results for the conserved quantities for the Lifshitz black holes in CG in the presence of a conformally invariant scalar field, which satisfy the first law of thermodynamics. On the other hand, the entropy is computed using Wald's formula \cite{Wald:1993nt}.

The paper is organized as follows: In section II we study analytical AdS black holes and Lifshitz black holes to CG with a conformally invariant scalar field. In Section III we study the thermodynamical properties: we compute the Wald entropy, the Hawking temperature and the mass of the solutions; in addition, we study the horizon thermodynamics of the Lifshitz black holes with $z=4$. We conclude with final remarks in Section IV.


\section{AdS and Lifshitz black hole solutions}
In this section we present AdS black hole solutions and Lifshitz black hole solutions of CG with a self-interacting conformally coupled scalar field. The Lifshitz black holes are characterized by dynamical exponents $z=0$ and $z=4$, while the AdS black holes correspond to the case $z=1$. The action of CG with a conformally-invariant scalar field is given by
\begin{equation}\label{action}
I=\int d^4 x \sqrt{-g} \left[ \frac{1}{2 \alpha} C^{\mu \nu \rho \sigma} C_{\mu \nu \rho \sigma} -  \left(  \frac{1}{2} g^{\mu \nu} \nabla_{\mu} \phi \nabla_{\nu} \phi +\frac{1}{12} \phi^2 R +\nu \phi^4 \right) \right] \, ,
\end{equation}
where $C_{\mu \nu \rho \sigma}$ is the Weyl tensor, which is the totally traceless part of the Riemann tensor, $R$ is the Ricci scalar, $\phi$ is the scalar field, $\alpha$ is a dimensionless gravitational coupling constant and $\nu$ is the dimensionless coupling constant of the potential $\phi^4$. Varying the action with respect to the metric and with respect to the scalar field, the following field equations are obtained:
\begin{equation}
B_{\mu \nu} \equiv 2 \nabla^{\rho} \nabla^{\sigma} C_{\mu \rho \sigma \nu}+ R^{\rho \sigma} C_{\mu \rho \sigma \nu}=- \frac{\alpha}{2} T_{\mu \nu} \, ,
\end{equation}
where $B_{\mu \nu}$ is the Bach tensor and the energy-momentum tensor is given by
\begin{equation}
T_{\mu \nu}= \partial_{\mu} \phi \partial _{\nu} \phi -g_{\mu \nu}\left( \frac{1}{2} \nabla_{\alpha} \phi \nabla ^{\alpha} \phi + \nu \phi ^4 \right) +\frac{1}{6} \left(  g_{\mu \nu} \nabla_{\alpha} \nabla ^{\alpha} -\nabla_{\mu} \nabla_{\nu}  +G_{\mu \nu} \right) \phi^2 \,,
\end{equation}
and the Klein-Gordon equation reads
\begin{equation}
\nabla_{\alpha} \nabla^{\alpha} \phi -4 \nu \phi ^3-\frac{R}{6} \phi =0 \, .
\end{equation}

Now, in order to solve the field equations, we will consider the following ansatz for the Lifshitz black hole metric
\begin{equation}\label{metric}
ds^2=-\left( \frac{r}{\ell}  \right)^{2z} f(r) dt^2+\frac{\ell^2}{r^2 f(r)} dr^2+ r^2 d \Omega_k^2 \,,
\end{equation}
where
\begin{equation}
d \Omega^2_k= \left\{ \begin{array}{ccc}
d\theta^2+ \sin^2 \theta d\phi & \,\,\, k=1 \\
d\theta^2 + d\phi^2 & \,\,\, k=0 \\
d\theta^2+ \sinh^2 \theta d\phi & \,\,\, k=-1 \\
\end{array}
\right.
\end{equation}
and where $k$ parameterizes the curvature of the spatial transverse section; thus, with this ansatz we permit the possibility of black holes with spherical, flat and hyperbolic event horizons. Also, in order to the metric ($\ref{metric}$) tends asymptotically to the Lifshitz space-time, it is required that $f(r) \rightarrow 1$ when $r \rightarrow \infty$. On the other hand, we will consider a scalar field regular outside and on the event horizon which only depends on the radial coordinate $\phi=\phi (r)$. We obtain analytical solutions for $z=0$, $z=1$ and $z=4$, which are given by:

\begin{itemize}

\item Solution with null dynamical exponent $z=0$:
\begin{equation}
\phi (r)= \frac{\phi_0}{r} \,,
\end{equation}
\begin{equation}\label{metric1}
f(r)=1+\frac{\ell^2(k+12 \nu \phi_0^2)}{4 r^2}\,,
\end{equation}
where $\phi_0$ is an integration constant, and the parameters must satisfy the condition $\phi_0^2 (k+6 \nu \phi_0^2) (\alpha +96 \nu)=0$. Notice that in the case $k+6 \nu \phi_0^2=0$, the single integration constant is fixed to $\phi_0 ^2 =-k/( 6\nu )$. In what follows we will focus our attention to the case $\alpha +96 \nu=0$. When $k+12 \nu \phi_0^2 <0 $ there is an event horizon located at $r_+=\frac{1}{2} \sqrt{-\ell^2(k+12 \nu \phi_0^2)}$.

\item Solution with dynamical exponent $z=1$, which corresponds to asymptotically AdS black holes:
\begin{equation}
\phi (r)= \frac{\phi_0}{r} \,,
\end{equation}
\begin{equation}\label{metric2}
f(r)=1+\frac{A}{r}+\frac{\ell^2(k+12 \nu \phi_0^2)}{r^2}\,,
\end{equation}
where $A$ and $\phi_0$ are integration constants, and the parameters must satisfy the condition $\phi_0^2 (k+6 \nu \phi_0^2) (\alpha +96 \nu)=0$. An event horizon can be located at $\frac{1}{2} \left( -A+ \sqrt{A^2 -4\ell^2 (k+12 \nu \phi_0^2)} \right)$ for some conditions on the parameters. For $A \geq 0$, then $4 \ell^2 (k+12 \nu \phi_0^2)<0$ is required for the existence of the event horizon, while for $A<0$, it is required $A^2> 4 \ell^2 (k+12 \nu \phi_0^2)$. 
As before, in the case $k+6 \nu \phi_0^2=0$ the integration constant $\phi_0$ is fixed to $\phi_0 ^2 =-k/( 6\nu )$. In what follows we will focus our attention to the case $\alpha +96 \nu=0$.

\item Solution with dynamical exponent $z=4$ :

\begin{equation}
\phi (r)= \frac{\phi_0}{r^3} \,,
\end{equation}
\begin{equation}\label{metric3}
f(r)=1+\frac{k \ell^2}{4 r^2}+\frac{\nu \ell^2 \phi_0^2}{2 r^6} \,,
\end{equation}
where $\phi_0$ is an integration constant.  The analytical expression for the event horizon is more complicated for this metric, and we will not show it.

\end{itemize}

As final comment, the curvature invariants $R$, $R_{\mu \nu} R^{\mu \nu}$ and $R_{\mu \nu \rho \sigma} R^{\mu \nu \rho \sigma}$ of the solutions presented here are all regular outside and on the event horizon and they are singular at $r=0$. However, it is worth pointing out that the curvature scalars, such as the Kretschmann scalar $R_{\mu \nu \rho \sigma} R^{\mu \nu \rho \sigma}$, which are invariant under general coordinate transformations, are not invariant under conformal transformations of the metric tensor.  Therefore, the conformal invariance can solve the problem of space-time singularities \cite{narlikar:1977nf}. In \cite{Bambi:2016wdn}, suitable conformal factors were found that make the solution conformally equivalent to the Schwarzschild solution regular everywhere.

\section{Thermodynamics}

In this section we will determine the thermodynamical properties of the black hole solutions presented in the previous section, such as the mass, the Wald entropy and the Hawking temperature, and we will show that they satisfy the first law of black hole thermodynamics.

 \subsection{Entropy and temperature}
 
 In this section we shall compute the Wald entropy to the black hole solutions, which is given by
 \begin{equation}
 S=-2 \pi \int_{\Sigma} \left( \frac{\partial L}{\partial R_{\mu \nu \rho \sigma}}  \right) \hat{\epsilon}_{\mu \nu } \hat{\epsilon }_{\rho \sigma}  \sqrt{h} d\theta d\phi \, ,
 \end{equation}
 where $\hat{\epsilon}_{\mu \nu}$ is the binormal vector to the bifurcation surface $\Sigma$ and it is normalized to $\hat{\epsilon}_{\mu \nu} \hat{ \epsilon}^{\mu \nu}=-2$, $h$ is the metric determinant on the surface and
 \begin{equation}
 P^{\mu \nu \rho \sigma} \equiv \frac{\partial L}{\partial R_{\mu \nu \rho \sigma}}=\frac{1}{\alpha}C^{\mu \nu \rho \sigma}-\frac{1}{24} \phi^2 (g^{\mu \rho}g^{\nu \sigma}-g^{\mu \sigma}g^{\nu \rho})\, .
 \end{equation}
 On the other hand, the Hawking temperature of the Lifshitz black holes can be found by evaluating the surface gravity $\kappa$, which is defined by $\kappa^2=-\frac{1}{2}(\nabla_{\mu} \chi_{\nu})(\nabla^{\mu} \chi^{\nu})$, where $\chi$ is the time-like Killing vector $\chi^{\nu}=(1,0,0,0)$; thus, for the metric (\ref{metric}) we obtain $\kappa=\frac{1}{2}  \left(\frac{r_+}{\ell} \right)^{1+z}  f'(r_+)$ and
 \begin{equation}
 T=\frac{\left(\frac{r_+}{\ell} \right)^{1+z} f'(r_+)}{4 \pi} \, .
 \end{equation}


 The Wald entropy of the Lifshitz black holes with $z=4$ is given by
 
 \begin{eqnarray}\label{entropia}
 \notag S &=& -8 \pi r_+^2 \left( \frac{\phi_0^2}{24 r_+^6} +\frac{4}{\alpha \ell^2} \right) \omega_{k} \\
 &=& \left( \frac{2 \pi r_+^2 ( \alpha -48 \nu )}{ 3 \alpha \nu \ell^2} +\frac{  k \pi}{6 \nu}  \right) \omega_k \, ,
 \end{eqnarray}
where $\omega_k$ denotes the volume of $d \Omega_k$ . In this case, the positivity of the entropy requires that $\alpha <0$. On the other hand, the Hawking temperature of the Lifshitz black holes with $z=4$ is given by
 \begin{equation}\label{temperatura}
 T=\frac{r_+^2}{4 \pi \ell^3} \left(  \frac{6 r_+^2}{\ell^2}+k \right) \, .
 \end{equation}

 For the AdS black holes and the Lifshitz black holes with $z=0$, the following expression for the Wald entropy is obtained
 \begin{equation}
 S=-\frac{ \pi \omega_{k} (\alpha+96 \nu) \phi_0^{2}}{3 \alpha} = 0 \, .
 \end{equation} 
where, by imposing the condition $\alpha +96 \nu=0$, it gives a null entropy. In the next section, we will show that these solutions also have zero mass, and therefore the first law of black hole thermodynamics is trivially satisfied. The Hawking temperature of the Lifshitz black holes with $z=0$ is given by
   \begin{equation}
 T=\frac{1}{2 \pi \ell} \, ,
 \end{equation}
notice that the same temperature is obtained for the Lifshitz black hole with $z=0$ in three-dimensional conformal gravity \cite{Gonzalez:2017ptj}, while the temperature of the AdS black holes is found to be
     \begin{equation}
 T=\frac{1}{4 \pi} \left( \frac{r_+}{\ell^2}-\frac{k+12 \nu \phi_0^2}{r_+} \right) \, .
 \end{equation}
 
\subsection{Conserved charges}

In order to compute the conserved charges we shall employ the generalized ADT formalism proposed in \cite{Kim:2013zha}, which corresponds to the off-shell generalization of the on-shell Noether potential of the ADT formalism \cite{Abbott:1981ff, Deser:2002rt}. Also, this formalism was implemented with a one parameter path in the solution space, such as in \cite{Barnich:2001jy, Barnich:2003xg}. See also the Refs. \cite{Peng:2014gha, Peng:2015yjx} where the off-shell Noether current was constructed using the variation of the Bianchi identity. In those references the method was applied to compute the mass of some black hole solutions in conformal Weyl gravity, in Einstein Gauss-Bonnet gravity and in Horndeski  theory. This method have been used also to compute the conserved charges of asymptotically Lifshitz black holes of gravity theories whose action contains higher order curvature terms \cite{Gim:2014nba}, and to compute the mass of three-dimensional Lifshitz black hole solutions of New Massive Gravity with a non-minimally coupled scalar field \cite{Ayon-Beato:2015jga}. Here, we shall apply the method to CG with scale invariant matter fields.

 Now, we will review briefly the method. The variation of the action (\ref{action}) yields
\begin{equation}
\delta I= \int d^4x   \sqrt{-g} \left( \mathcal{E} _{\mu \nu} \delta g^{\mu \nu} + \mathcal{E} _{ (\phi)} \delta \phi + \nabla_{\mu} \Theta ^{\mu} (g; \delta g) \right)\,,
\end{equation}
where $\mathcal{E}_{\mu \nu}=0$ and $\mathcal{E}_ { (\phi )}=0$ denotes the equations of motion (EOM) for the metric and for the scalar field, respectively, and $\Theta ^{\mu} (g; \delta g)$ is a surface term.

By equating a variation induced by a diffeomorphism generated by a smooth vector field $\xi$ with the generic variation, an off-shell conserved current can be obtained \cite{Deruelle:2003ps, Padmanabhan:2009vy}
\begin{equation}
J^{\mu} (g; \xi)= \frac{L}{\sqrt{-g}} \xi ^{\mu} +2 \mathcal{E}^{\mu \nu} \xi_{\nu}-\Theta^{\mu} (\mathcal{L}_{\xi}g, \mathcal{L}_{\xi}\phi)\,,
\end{equation}
where $\mathcal{L}_{\xi}$ denotes the Lie derivative with respect to the vector field $\xi $. From this current an off-shell Noether potential $K^{\mu \nu}$ can be defined by $J^{\mu}=\nabla_{\nu} K^{\mu \nu}$. On the other hand, an off-shell Noether current can be defined as \cite{Bouchareb:2007yx}
\begin{equation}
J^{\mu}_{ADT}=\delta \mathcal{E}^{\mu \nu} \xi_{\nu} + \mathcal{E}^{\mu \alpha} h_{\alpha \nu} \xi^{\nu}-\frac{1}{2} \xi ^{\mu} \mathcal{E}^{\alpha \beta} h_{\alpha \beta}+\frac{1}{2} \mathcal{E}^{\mu}_{\,\, \nu} \xi^{\nu} h \,,
\end{equation}
where $h_{\mu \nu}$ denotes the variation of the background metric
\begin{equation}
h_{\mu \nu}= \delta g_{\mu \nu} \,,
\end{equation}
and the indices are raised an lowered with the background metric $g^{\mu \nu}$ or $g_{\mu \nu}$, and $h=g^{\mu \nu} h_{\mu \nu} $. The off-shell ADT potential $Q^{\mu \nu}_{ADT}$ is defined by $J^{\mu}_{ADT}=\nabla_{\nu} Q^{\mu \nu}_{ADT}$. 
A very important relation between the off-shell Noether potential and the ADT potential is:
\begin{equation}
\sqrt{-g} Q^{\mu \nu}_{ADT}= \frac{1}{2} \delta (\sqrt{-g}K^{\mu \nu})-\sqrt{-g} \xi^{[\mu}\Theta^{\nu]} (g; \delta g) \,.
\end{equation}

Using this Noether potential, one can define the quasi-local conserved charge by
\begin{eqnarray}\label{masa}
\notag \mathcal{Q}(\xi) &=& 2 \int_{0}^{1}ds \int d^ 2 x_{\mu \nu} \sqrt{-g} Q^{\mu \nu}_{ADT} (g; s) \\
 &=& \int d^ 2 x_{\mu \nu} \Delta \hat{K}^{\mu \nu}- 2 \int_{0}^{1} ds \int \sqrt{-g} d^ 2 x_{\mu \nu} \xi^{[\mu } \Theta ^{\nu ]} (g; s) \,,
\end{eqnarray}
 where $d^2 x_{\mu \nu}=\frac{1}{4 \sqrt{-g}} \epsilon_{\mu \nu \rho \sigma} dx^{\rho} \wedge dx^{\sigma}$ and
 \begin{equation}
 \Delta \hat{K} = \sqrt{-g} K^{\mu \nu} \Big|_{s=1}- \sqrt{-g} K^{\mu \nu} \Big|_{s=0} \, .
 \end{equation}
The integration is performed along a path with parameter $s$ in the solution space $0 \leq s \leq 1$, and a free parameter $\mathcal{Q}$ in the solutions of EOM is parametrized as $ s \mathcal{Q} $.
 Eq. (\ref{masa}) is the main result that will allows us to evaluate the conserved charges (mass) of the black hole solutions.

The boundary term can be written as
\begin{equation}
\Theta^{\mu} (g, \delta g)= \Theta^{\mu}_{W}+ \Theta^{\mu}_{(\phi)} \,,
\end{equation}
where the first term $\Theta^{\mu}_{W}$ denotes the contribution from  the Weyl-squared term and the second term $\Theta^{\mu}_{(\phi)}$ denotes the contribution coming from the scalar field. The generic formula for the surface term in theories which contains only invariants of the curvature tensor can be taken by \cite{Kim:2013zha}
 \begin{equation}\label{tw}
 \Theta^{\mu}_{W} (\delta g)= 2 [ P^{\mu (\alpha \beta) \gamma} \nabla_{\gamma} \delta g_{\alpha \beta}-\delta g_{\alpha \beta} \nabla _{\gamma} P^{\mu (\alpha \beta) \gamma}] \,,
 \end{equation}
 where 
 \begin{equation}
P^{\mu \nu \rho \sigma}= \frac{ \partial L}{\partial R_{\mu \nu \rho \sigma}} \,.
\end{equation}
On the other hand, we have found that the boundary term coming from the self-interacting conformally coupled scalar field reads 
 \begin{equation}\label{tph}
 \Theta^{\mu}_{(\phi) }=-\delta \phi \nabla^{\mu} \phi-\frac{1}{12} \left( \phi^2 \nabla_{\nu} h^{\mu \nu} -\phi^2 \nabla^{\mu} h +h \nabla^{\mu} \phi^2 -h ^{\mu \nu} \nabla_{\nu } \phi^2 \right) \,.
 \end{equation}
 
 Using Eqs. (\ref{tw}) and (\ref{tph}), the Noether potentials are found to be
 
 \begin{eqnarray}
 K^{\mu \nu}_W &=& \frac{2}{\alpha}\left(  C^{\mu \nu \rho \sigma} \nabla_{ [ \rho} \xi_{\sigma ]} -2 \xi_{\sigma}  (\nabla^{[ \mu} R^{\nu ] \sigma}+\frac{1}{6} g^{\sigma [ \mu} \nabla ^{ \nu ] } R) \right) \\
 K^{\mu \nu}_{(\phi)} &=& \frac{1}{6} \left( \frac{1}{2} \phi^2 \nabla ^{\nu} \xi^{\mu} -\frac{1}{2} \phi^2 \nabla^{\mu} \xi^{\nu} +\xi^{\nu} \nabla^{\mu} \phi^2 -\xi^{\mu} \nabla ^{\nu} \phi^2 \right)\,.
 \end{eqnarray}

Using expression (\ref{masa}), we shall compute the masses associated to our solutions. For the time-like Killing vector $\xi = -\partial_{t}$ the charge (\ref{masa}) corresponds to the mass. 

Now, we consider the Lifshitz black holes with $z=4$ and we take an infinitesimal parametrization of a single parameter path, by considering the perturbation
\begin{equation}
\phi_0 \rightarrow \phi_0 + d\phi_0 \, ,
\end{equation}
the Noether potentials related to the energy reads

\begin{eqnarray}
\sqrt{-g}K^{t r} &=& -\frac{16 r^6 \sigma_k(\theta)}{ \alpha \ell^7}- \frac{(5\alpha -48 \nu) \phi_0^2 \sigma_k (\theta)}{3 \alpha \ell^5} + \mathcal{O} (\frac{1}{r^2}) \\
\sqrt{-g}Q^{tr} &=& -\frac{(\alpha-48 \nu) \phi_0 d\phi_0 \sigma_k (\theta)}{6 \alpha \ell^5} \, ,
\end{eqnarray}
where $\sigma_k(\theta)= \sin \theta, 1, \sinh \theta$ for $k=1,0, -1$ respectively. So, the mass of the Lifshitz black holes with $z=4$ is given by
 
 \begin{eqnarray}
 \notag \mathcal{M} &=& -\frac{(\alpha-48\nu) \phi_0^2 \omega_{k}}{ 6 \alpha \ell^5} \\
 &=& (\alpha-48 \nu) \left( \frac{r_{+}^6}{3 \alpha \nu \ell^7} + \frac{  k r_+^4 }{12 \alpha \nu \ell^5}  \right) \omega_k \, .
 \end{eqnarray}
 
 Using this result for the mass together with the entropy given in Eq. (\ref{entropia}) and the temperature given in Eq. (\ref{temperatura}), it is straightforward to show that the first law of black hole thermodynamics $d \mathcal{M} = T dS$ holds.

For the asymptotically AdS black holes and the Lifshitz black holes with $z=0$, we find that they have zero mass $\mathcal{M}=0$ in all the cases, and this result does not depend of the values of $k$. Also, as we found in the previous section, they have zero entropy, therefore there is not a conserved charge associated to the integration constants, which can be interpreted as a kind of gravitational hair. Black hole solutions with planar horizon having zero mass and zero entropy have been reported in Lovelock gravity with non-minimally coupled scalar fields \cite{Correa:2013bza} and also in the three-dimensional New Massive Gravity with non-minimally coupled scalar fields \cite{Correa:2014ika}. The solutions presented in those papers depends on a single integration constant which was interpreted as a gravitational hair, due to there is not a conserved charge associated to it. On the other hand, in pure Lovelock gravity black holes with gravitational hair have been found \cite{Anabalon:2011bw}.

\subsection{Horizon thermodynamics}

In this section, we will employ the method developed by Padmanabhan in \cite{Padmanabhan:2012gx} in order to study the conserved quantities of the Lifshitz black holes with dynamical exponent $z=4$, which, as we have seen, are the only solutions with non-trivial thermodynamics.  The method is based on the relation between the first law of thermodynamics and the equations of motion, and it seems to be applicable to a very wide class of theories. In this approach, the conserved quantities can be obtained directly from the equations of motion evaluated on the event horizon. So, considering the $\mathcal{E}^r_{\,\, r}$ component of the field equations, and evaluating the second and third derivatives of $f(r)$ at the horizon $r_+$, we arrive to the following equation:
\begin{equation}
\frac{1}{48 r_+^{12} \ell^4} \left( (-k r_+^4-6 \nu \phi_0 ^2 ) \ell ^2-2 r_+^7 f'(r_+) \right) \left((-7  k r_+^4 +2 (4 \alpha +147 \nu ) \phi_0^2) \ell^2 +14 r_+^7 f'(r_+)\right)=0~. 
\end{equation}
Now, considering that only the first factor in parenthesis can be zero, this equation reduces to
\begin{equation}
 (-k r_+^4-6 \nu \phi_0 ^2 ) \ell ^2-2 r_+^7 f'(r_+) =0\,.
\end{equation}
Writing $\phi_0$ in terms of the horizon radius $\phi_0^2=-(k+\frac{4 r_+^2}{\ell^2}) \frac{r_+^4}{2 \nu}$, and multiplying the above equation by the factor $\frac{(\alpha-48 \nu) \omega_k dr_+}{6 \alpha \ell^2 \nu r_+}$ we obtain
\begin{equation}
d \left[  \left(\alpha- 48 \nu\right) \left(  \frac{r_+^6}{3 \alpha \nu \ell^7} + \frac{k r_+^4}{12 \alpha \nu \ell^5}\right) \omega_{k}  \right] = \frac{1}{4 \pi} \left(  \frac{r_+ }{ \ell} \right) ^5 f'(r_+) d \left( \frac{2 \pi r_+^2 (\alpha- 48 \nu) \omega_{k} }{3 \alpha \nu \ell^2 } + a \right) \,,
\end{equation}
where $a$ is an additive constant. Therefore, we arrive to an equation with the same form of the first law of black hole thermodynamics. Finally, identifying the Hawking temperature and choosing the constant as $a=\frac{k \pi}{6 \nu}$, the conserved quantities coincide with the mass and entropy obtained before.

\section{Quasinormal modes}
In this section we will study the propagation of a conformally coupled scalar field in the backgrounds found in Section II, and then we find the quasinormal modes analytically in some cases. Perturbations of a conformally coupled test scalar field obey the general relativistic Klein-Gordon equation
\begin{equation}
\frac{1}{\sqrt{-g}}\partial _{\mu }\left( \sqrt{-g}g^{\mu \nu }\partial
_{\nu } \varphi \right) =\frac{R}{6}\varphi \,, \label{KGNM}
\end{equation}%
where $R$ is the Ricci escalar. With the help of the following ansatz
$\varphi =e^{-i\omega t}Y(\Omega)R(r),$
the Klein-Gordon equation reduces to the form
\begin{eqnarray}
\notag && \frac{d^2 R(r)}{dr^2}+\left(\frac{3+z}{r}+\frac{f'(r)}{f(r)}\right) \frac{dR(r)}{dr} + \frac{\ell^2}{6 r^2 f(r)}\Big(\frac{6 \omega^2 \left( r/\ell \right)^{-2z}}{f(r)}-\frac{2 \left(k+3\kappa^2  \right)}{r^2} \\
&& +\frac{2(3+z(2+z))f(r)+r((5+3z)f'(r)+rf''(r))}{\ell^2}     \Big)R(r)=0~,
\label{radial}
\end{eqnarray}
where $-\kappa^2$ is the eigenvalue of the Laplacian in the base submanifold; so, for spherical horizon $\kappa^2=l(l+1)$ where $l=0,1,2,...$, for flat horizon $\kappa^2$ is a non-negative integer and for hyperbolic horizon $\kappa^2=1/4 + \xi^2$, where $\xi \geq 0$ takes discrete real values \cite{Balazs:1986uj}. Now, defining $R(r)$ as
$ R(r)=\tilde{R}(r) / r $
and using the tortoise coordinate $r_*$ given by
$ dr_* = \ell^{z+1} dr/ (r^{z+1}f(r) )$,
 the Klein-Gordon equation can be written as a one-dimensional Schr\"{o}dinger equation,
 \begin{equation}\label{ggg}
 -\frac{d^{2}\tilde{R} (r_*)}{dr_*^{2}}+V(r)\tilde{R}(r_*)=\omega^{2}\tilde{R}(r_*)\,,
 \end{equation}
 with an effective potential $V(r)$, which is parametrically thought of as $V(r_*)$, given by
  \begin{eqnarray}\label{pot}
\notag V(r) &=& - \Big(\frac{r}{\ell} \Big)^{2z+2} f(r)^2 \Big[\frac{2}{r^2}-\frac{1}{r} \Big(\frac{z+3}{r} +\frac{f'(r)}{f(r)} \Big)+\frac{1}{6 r^2 f(r)} \Big(-\frac{2(k+3 \kappa^2) \ell^2}{r^2} +2\left(3+z(z+2) \right)f(r)   \\
 && +r \left((3z+5)f'(r) +rf''(r) \right) \Big) \Big] ~.
 \end{eqnarray}


\subsection{Quasinormal modes for Lifshitz black holes with $z=0$}

In this section, we deduce the exact analytic solution for the Klein--Gordon equation (\ref{radial}) in the background described by (\ref{metric1}). Under the change of variable $y=1-\frac{r_+^2}{r^2}$, the equation (\ref{radial})  can be written as
\begin{equation}
R''(y)+\frac{1}{y} R'(y)+\frac{1}{12 y^2 (1-y)^2}\left( 3 \omega^2 \ell^2  +2y+y^2+\frac{4 (k+3 \kappa^2) y (1-y)}{k+12 \nu \phi_0^2}  \right)R(y)=0~,
\end{equation}
and if, in addition, we define $R(y)=y^{\alpha}(1-y)^\beta F(y)$, the above equation leads to the hypergeometric equation
\begin{equation}\label{hypergeometric}
 y(1-y)F''(y)+\left[c-(1+a+b)y\right]F'(y)-ab F(y)=0~,
\end{equation}
where
\begin{equation}
\alpha =  \pm\frac{i\omega\ell}{2}, \quad \beta= \frac{1}{2}\pm\frac{i\omega\ell}{2} ~,
\end{equation}
and the constants are given by
\begin{equation}\label{a}
a_{1,2}= \alpha +\beta \pm \frac{\sqrt{(k+4 \kappa^2 -4 \nu \phi_0^2) (k+12 \nu \phi_0^2)}}{ 2 (k+12 \nu \phi_0^2) } ~,
\end{equation}
\begin{equation}
b_{1,2}= \alpha +\beta \mp \frac{\sqrt{(k+4 \kappa^2 -4 \nu \phi_0^2) (k+12 \nu \phi_0^2)}}{ 2 (k+12 \nu \phi_0^2) } ~.
\end{equation}
\begin{equation}
c=1+2\alpha~.
\end{equation}
The general solution of the hypergeometric equation~(\ref{hypergeometric}) is
\begin{equation}
F(y)= C_{1}\,\, {_2}F{_1}(a,b,c;y)+C_2y^{1-c}\,{_2}F{_1}(a-c+1,b-c+1,2-c;y)~,
\label{HSolution}
\end{equation}
and it has three regular singular points at $y=0$, $y=1$, and
$y=\infty$. Here, ${_2}F{_1}(a,b,c;y)$ is a hypergeometric function,
and $C_{1}$ and $C_{2}$ are integration constants.
Thus, in the vicinity of the horizon $y=0$ and using
the property $F(a,b,c;0)=1$, the function $R(y)$ behaves as
\begin{equation}\label{Rhorizon}
R(y)=C_1 e^{\alpha \ln y}+C_2 e^{-\alpha \ln y} ~,
\end{equation}
so that the scalar field $\varphi$, for $\alpha=\alpha_-$, can be written as follows:
\begin{equation}
\varphi\sim C_1 e^{-i\omega (t+ (\ell \ln y )/2)}+C_2e^{-i\omega (t-(\ell \ln y )/2)}~.
\end{equation}
Here, the first term represents an ingoing wave, and the second represents an outgoing wave near the black hole horizon. Imposing the requirement of
only ingoing waves on the event horizon, we fix $C_2=0$. Then, the radial solution can be written as
\begin{equation}\label{horizonsolution}
R(y)=C_1 e^{\alpha \ln y}(1-y)^\beta {_2}F{_1}(a,b,c;y)= C_1e^{-\frac{i\omega \ell}{2} \ln y}(1-y)^\beta{_2}F{_1}(a,b,c;y)~.
\end{equation}
To implement boundary conditions at infinity ($y=1$), we apply Kummer's formula for the hypergeometric function \cite{M. Abramowitz},
\begin{equation}\label{relation}
{_2}F{_1}(a,b,c;y)=\frac{\Gamma(c)\Gamma(c-a-b)}{\Gamma(c-a)\Gamma(c-b)}{_2}F{_1}(a,b,a+b-c;1-y)+(1-y)^{c-a-b}\frac{\Gamma(c)\Gamma(a+b-c)}{\Gamma(a)\Gamma(b)}{_2}F{_1}(c-a,c-b,c-a-b+1;1-y) ~.
\end{equation}
Taking into consideration the above expression, the radial function (\ref{horizonsolution}) reads
\begin{eqnarray}
\notag R(y) &=& C_1 e^{-\frac{i\omega \ell}{2} \ln y}(1-y)^\beta\frac{\Gamma(c)\Gamma(c-a-b)}{\Gamma(c-a)\Gamma(c-b)} {_2}F{_1}(a,b,a+b-c;1-y)  \\
  && + C_1 e^{-\frac{i\omega \ell}{2}  \ln y}(1-y)^{1-\beta}\frac{\Gamma(c)\Gamma(a+b-c)}{\Gamma(a)\Gamma(b)} {_2}F{_1}(c-a,c-b,c-a-b+1;1-y)\,,
\end{eqnarray}
and at infinity, it can be written as
\begin{equation}\label{R2}\
R_{asymp.}(y) = C_1 (1-y)^\beta \frac{\Gamma(c)\Gamma(c-a-b)}{\Gamma(c-a)\Gamma(c-b)}+ C_1 (1-y)^{1-\beta}\frac{\Gamma(c)\Gamma(a+b-c)}{\Gamma(a)\Gamma(b)}~.
\end{equation}
The effective potential (\ref{pot}) tends to zero at infinity; thus, considering $\beta=\beta_+$ and imposing as a boundary condition that only outgoing waves exist at spatial infinity, we set $c-a=-n$ or $c-b=-n$ for $n=0,1,2,...$. Therefore, the QNFs for the Lifshitz black holes with dynamical exponent $z=0$ are given by
\begin{equation}\label{w1}
\omega = \pm \frac{1}{2 \ell} \sqrt{-\frac{k+4 \kappa^2-4 \nu \phi_0^2}{k+12 \nu \phi_0^2}} -\frac{i}{2 \ell} (1+2n) ~.
\end{equation}
The imaginary part of the QNFs is always negative for $k=1,0,-1$ and $\nu<0$; so, in these cases the propagation of a scalar field is formally stable in this background. However, for the topological black holes with $k=-1$ and $\nu > \xi^2 /\phi_0^2$ the QNFs are purely imaginary, and if the parameter $\xi^2$ takes values in the range $0 \leq \xi^2<4\nu \phi_0^2-1/4$, the imaginary part of the fundamental quasinormal frequency is positive, implying that the propagation of a scalar field on the topological black hole is unstable.

\subsection{Quasinormal modes for AdS black holes}

Under the change of variable $y=\frac{r-r_{+}}{r-r_{-}}$, the Klein-Gordon equation (\ref{radial}) in the background described by ($\ref{metric2}$) can be written as

\begin{eqnarray}
\notag  &&R''(y)+\left( \frac{1}{y}-\frac{2}{y-1}+\frac{2}{y-y_0} \right)R'(y)+\frac{1}{y(y-1)(y-y_0)}\Big( \frac{y_0^2 \omega^2 \ell^2}{y(y_0-1)^2 (k+12 \nu \phi_0^2)}-\frac{2(y_0-1)}{y-1} +\frac{y_0 (y_0-1)(-\kappa^2+4 \nu \phi_0^2)}{(y-y_0)(k+12 \nu \phi_0^2)}   \\
&&     +\frac{y y_0 \omega^2 \ell^2}{(y_0-1)^2 (k+12 \nu \phi_0^2)} +\frac{(y_0-1)^2 (k-ky_0-y_0 \kappa^2)-y_0 (1+y_0) \omega^2 \ell^2 -4 (y_0-1)^2 (-3+2y_0) \nu \phi_0^2}{(y_0-1)^2 (k+12 \nu \phi_0^2)} \Big) R(y)=0   ~,
\end{eqnarray}
where $y_0=\frac{r_{+}}{r_{-}}$. If, in addition, we define $R(y)=y^{\alpha} (1-y)^{\beta}(y_0-y)^{\gamma} H(y)$, the above equation leads to the general Heun equation

\begin{equation}
H''(y)+\left( \frac{\mu}{y}+\frac{\delta}{y-1}+\frac{\epsilon}{y-y_0} \right)H'(y)+\frac{\lambda \xi y- \eta}{y(y-1)(y-y_0)} H(y)=0 ~,
\end{equation}

where
\begin{equation}
\alpha=\pm \frac{i \omega \ell}{y_0-1} \sqrt{\frac{y_0}{k+12 \nu \phi_0^2}}\,, \,\,\,\, \beta=\frac{3}{2} \pm \frac{1}{2}\,, \,\,\, \gamma=-\frac{1}{2}\pm \frac{1}{2}\sqrt{\frac{k+4 \kappa^2-4 \nu \phi_0^2}{k+12 \nu \phi_0^2}} ~,
\end{equation}
and the constants are given by

\begin{eqnarray}
\notag  &&  \mu=1+2\alpha\,, \,\,\, \delta=2(-1+\beta)\,, \,\,\, \epsilon= 2(1+\gamma) \\
&& \lambda = 2 \alpha+\beta+\gamma \,, \,\,\, \xi= \beta+ \gamma \\
\notag && \eta=-1+\beta (3-\beta)+2 \alpha (1 +  \gamma)+\gamma (2+ \gamma ) +\frac{y_0}{3}  \left( 2+ 6\alpha (\beta-1) +3 \beta (\beta-2) -3 \gamma (\gamma+1) +\frac{k+3 \kappa^2}{k+12 \nu \phi_0^2} \right) ~,
\end{eqnarray}
with $\epsilon=\lambda+\xi-\mu-\delta+1$.
The properties of the general Heun function are not well known yet, and therefore we cannot extract the QNMs analytically in the general case. However, there are some special cases in which the Heun function can be reduced to the confluent Heun function or to the hypergeometric function, whose properties are better known. We will focus our attention to these cases in the following.

\subsubsection{$k+12 \nu \phi_0^2=0$}

First, we consider the special case $k+12 \nu \phi_0^2=0$. In this case there is only one horizon $r_{+}=-A$ for $A$ negative. Using the change of variable $y=1-\frac{r_{+}}{r}$, and defining $R(y)=y^{\alpha}(1-y)^{\beta} H(y)$, the Klein-Gordon equation leads to the confluent Heun differential equation
\begin{equation}\label{HC}
y(y-1)H''(y)+\left[ (b+1) (y-1) +(c+1) y \right] H'(y)+(dy- \epsilon) H(y)=0 ~,
\end{equation}
where
\begin{eqnarray}
\notag \alpha &=&\pm \frac{i \omega \ell^2}{r_{+}}\,, \,\,\, \beta=\frac{3}{2} \pm \frac{1}{2} ~,\\
b &=& 2 \alpha \,, \,\,\, c=2 \beta-3 \,,   \,\,\, d=-\frac{\ell^2}{3 r_+^2} (k+3\kappa^2) \,,
\epsilon=-(\alpha+\beta-1)^2-\frac{\ell^2}{3 r_{+}^2} (k+3\kappa^2+3 \omega^2 \ell^2) ~.
\end{eqnarray}
The solution of Eq. (\ref{HC}) is written as
\begin{equation}
H(y) = C_1 \text{Heun}_\text{C} \left( 0,b,c,d,-\frac{1}{2}(1+c)b-\frac{c}{2}-\epsilon,y \right)+C_2y^{-b} \text{Heun}_\text{C} \left( 0,-b,c,d,-\frac{1}{2}(1+c)b-\frac{c}{2}-\epsilon,y \right) ~,
\end{equation}

where Heun$_\text{C}$ is the confluent Heun function, and $C_1$ and $C_2$ are integration constants. So, the radial function $R(y)$ is given by

\begin{equation}
\notag R(y) = C_1 y^{\alpha} (1-y)^{\beta} \text{Heun}_\text{C} \left( 0,b,c,d,-\frac{1}{2}(1+c)b-\frac{c}{2}-\epsilon,y \right)+C_2y^{-\alpha} (1-y)^{\beta} \text{Heun}_\text{C} \left( 0,-b,c,d,-\frac{1}{2}(1+c)b-\frac{c}{2}-\epsilon,y \right) ~.
\end{equation}
Thus, in the vicinity of the horizon $y=0$ and using the property Heun$_\text{C} (0,b,c,d,e,0)=1$, the function $R(y)$ behaves as
\begin{equation}
R(y)=C_1 e^{\alpha \ln y}+C_2 e^{-\alpha \ln y},
\end{equation}
so that the scalar field $\varphi$, for $\alpha=\alpha_-$, can be written as follows:
\begin{equation}
\varphi\sim C_1 e^{-i\omega (t+ (\ell^2 \ln y )/r_+)}+C_2e^{-i\omega (t-(\ell^2 \ln y )/r_+)}~.
\end{equation}
Here, the first term represents an ingoing wave, and the second represents an outgoing wave near the black hole horizon. Imposing the requirement of
only ingoing waves on the event horizon, we fix $C_2=0$. Then, the radial solution can be written as
\begin{equation}\label{solution}
R(y)=C_1 e^{-\frac{i\omega \ell^2}{r_+} \ln y} (1-y)^{\beta} \text{Heun}_\text{C} \left( 0,b,c,d,-\frac{1}{2}(1+c)b-\frac{c}{2}-\epsilon,y \right)~.
\end{equation}

{\it Dirichlet boundary condition.} The effective potential (\ref{pot}) tends to $\left( \kappa^2-4 \nu \phi_0^2 \right)/ \ell^2$ and it is possible to impose as a boundary condition that the scalar field vanishes at spatial infinity. To implement boundary conditions at infinity ($y=1$), we use the  connection formula for the confluent Heun function \cite{aya, Kwon:2011ey}
\begin{eqnarray}
\notag \text{Heun}_{\text{C}}(0,b,c,d,e;y) &=&  D_1\frac{\Gamma(b+1)\Gamma(-c)}{\Gamma(1-c+k)\Gamma(b-k)}\text{Heun}_{\text{C}}(0,c,b,-d,e+d;1-y) \\
&& +D_2(1-y)^{-c}\frac{\Gamma(b+1)\Gamma(c)}{\Gamma(1+c+s)\Gamma(b-s)}\text{Heun}_{\text{C}}(0,-c,b,-d,e+d;1-y) ~,
\end{eqnarray}
where the parameters of the confluent Heun function satisfy the following equations
\begin{eqnarray}
&& k^2+(1-b-c)k-\epsilon-b-c+\frac{d}{2}=0 ~, \\
&& s^2+(1-b+c)s-\epsilon-b(c+1)+\frac{d}{2}=0 ~.
\end{eqnarray}
Taking into consideration the above expression, the radial function (\ref{solution}) at the asymptotic region $y \rightarrow  1$ reads
\begin{equation}\label{asy}\
R_{asymp.}(y) = C_1 D_1 (1-y)^\beta \frac{\Gamma(b+1)\Gamma(-c)}{\Gamma(1-c+k)\Gamma(b-k)}+ C_1D_2 (1-y)^{3-\beta}\frac{\Gamma(b+1)\Gamma(c)}{\Gamma(1+c+s)\Gamma(b-s)}~.
\end{equation}
Now, for $\beta=1$ or $\beta=2$, both terms of this expression go to zero for $y \rightarrow 1$. So, the Dirichlet boundary condition is automatically satisfied for any value of $\omega$, which implies a continuous spectrum and instability. It is worth mentioning that this kind of behavior have been observed in other AdS and Lifshitz black holes geometries for scalar type electromagnetic field perturbations \cite{Lopez-Ortega:2014oha, Gomez-Navarro:2017fyx}, see also Refs. \cite{ Estrada-Jimenez:2013lra, Cruz:2015nza} where continuous spectrum of frequencies have been obtained for a two-dimensional black hole. Nevertheless, it is expected that the QNFs depend on the physical parameters that describe the black hole and the test field. In \cite{Lopez-Ortega:2014oha, Gomez-Navarro:2017fyx} a different boundary condition was proposed, by canceling the leading term of the asymptotic behavior of the radial function well defined QNFs were obtained. We will show that for the AdS black hole a discrete set of QNFs, depending on the physical parameters of the black hole and the test field, can be obtained by considering Neumann boundary condition.

{\it Neumann boundary condition.} It is also possible to consider that the flux of the scalar field vanishes at spatial infinity, which implies the Neumann boundary condition.
The flux
\begin{equation} \label{flujo}
F=\frac{\sqrt{-g}g^{rr}}{2i} \left( R^{\ast} \partial_r R-R \partial_r R^{\ast} \right) ~,
\end{equation}
at infinity is given by
\begin{equation}
F(y \rightarrow 1) \propto -\frac{|C_1|^2 r_{+}^3 (3-2\beta)}{\ell^2} Im (A_1^{\ast}A_2) ~,
\end{equation}
where we have used the Eq. (\ref{asy}) to obtain the above expression, and $A_1$ and $A_2$ are given by
\begin{equation}
A_1=D_1 \frac{\Gamma(b+1)\Gamma(-c)}{\Gamma(1-c+k)\Gamma(b-k)}\, , \,\,\,\,\, A_2= D_2 \frac{\Gamma(b+1)\Gamma(c)}{\Gamma(1+c+s)\Gamma(b-s)} ~.
\end{equation}

The flux vanishes at infinity if $1-c+k=-n$, $b-k=-n$, $1+c+s=-n$ or $b-s=-n$ for $n=0,1,2,...$. Therefore, the discrete frequencies for the AdS black holes are given by
\begin{equation}
\omega=-i \frac{6(n+1)^2 r_+^2+\ell^2 (k+3 \kappa^2)}{12 \ell^2 (n+1) r_+}~,
\end{equation}
for $\beta=1$ or $\beta=2$.
The QNFs are purely imaginary and the imaginary part is always negative for $k=1,0$; so, in these cases the propagation of a scalar field is formally stable in this background. However, for the topological black holes with $k=-1$ and if the parameter $\xi^2$ takes values in the range $0 \leq \xi^2< 1/12-2(r_+ / \ell)^2$, the imaginary part of the fundamental quasinormal frequency is positive, implying that the propagation of a scalar field on the topological black hole is unstable.

\subsubsection{$A=0$}

In this case, under the change of variable $y=1-\frac{r_+^2}{r^2}$, the Klein--Gordon equation (\ref{radial})  can be written as
\begin{equation}
R''(y)+\left( \frac{1}{y}+\frac{1}{2(1-y)} \right) R'(y)+\frac{(y-1)(-\kappa^2 y+\omega^2 \ell^2)+2 k y+4 \nu \phi_0^2y (y+5) }{4 (k+12 \nu \phi_0^2) y^2 (1-y)^2}R(y)=0~,
\end{equation}
and if, in addition, we define $R(y)=y^{\alpha}(1-y)^\beta F(y)$, the above equation leads to the hypergeometric equation
\begin{equation}
 y(1-y)F''(y)+\left[c-(1+a+b)y\right]F'(y)-ab F(y)=0~,
 \label{hgeom}
\end{equation}
 where
\begin{equation}
\alpha =  \pm\frac{i\omega\ell}{2 \sqrt{-k-12\nu\phi_0^2}}, \quad \beta= \frac{3}{4}\pm\frac{1}{4} ~,
\end{equation}
and the constants are given by
\begin{equation}\label{a}
a_{1,2}= -\frac{1}{4}+\alpha +\beta \pm \frac{\sqrt{(k+4 \kappa^2 -4 \nu \phi_0^2) (k+12 \nu \phi_0^2)}}{ 4 (k+12 \nu \phi_0^2) } ~,
\end{equation}
\begin{equation}
b_{1,2}= -\frac{1}{4}+\alpha +\beta \mp \frac{\sqrt{(k+4 \kappa^2 -4 \nu \phi_0^2) (k+12 \nu \phi_0^2)}}{ 4 (k+12 \nu \phi_0^2) } ~.
\end{equation}
\begin{equation}
c=1+2\alpha~.
\end{equation}
The general solution of the hypergeometric equation~(\ref{hgeom}) is
\begin{equation}
F(y)= C_{1}\,\, {_2}F{_1}(a,b,c;y)+C_2y^{1-c}\,{_2}F{_1}(a-c+1,b-c+1,2-c;y)~,
\end{equation}
and it has three regular singular points at $y=0$, $y=1$, and
$y=\infty$. Here, ${_2}F{_1}(a,b,c;y)$ is a hypergeometric function,
and $C_{1}$ and $C_{2}$ are integration constants.
Thus, in the vicinity of the horizon $y=0$ and using
the property $F(a,b,c,0)=1$, the function $R(y)$ behaves as
\begin{equation}\label{Rhorizon}
R(y)=C_1 e^{\alpha \ln y}+C_2 e^{-\alpha \ln y} ~,
\end{equation}
so that the scalar field $\varphi$, for $\alpha=\alpha_-$, can be written as follows:
\begin{equation}
\varphi\sim C_1 e^{-i\omega \left(t+( \ell \ln y) /\left(2 \sqrt{-k-12 \nu \phi_0^2}\right)\right)}+C_2e^{-i\omega \left(t- (\ell \ln y ) /\left(2 \sqrt{-k-12 \nu \phi_0^2 }\right)\right)}~.
\end{equation}
Here, the first term represents an ingoing wave, and the second represents an outgoing wave near the black hole horizon. Imposing the requirement of
only ingoing waves on the event horizon, we fix $C_2=0$. Then, the radial solution can be written as
\begin{equation}\label{horizonsoluti}
R(y)=C_1 e^{\alpha \ln y}(1-y)^\beta {_2}F{_1}(a,b,c;y)= C_1e^{-(i\omega \ell \ln y)/\left(2 \sqrt{-k-12\nu \phi_0^2}\right)}(1-y)^\beta{_2}F{_1}(a,b,c;y)~.
\end{equation}

{\it Dirichlet boundary condition.} The effective potential (\ref{pot}) tends to $\left( \kappa^2-4 \nu \phi_0^2 \right)/ \ell^2$ and it is possible to impose as a boundary condition that the scalar field vanishes at spatial infinity. To implement boundary conditions at infinity ($y=1$), we apply Kummer's formula for the hypergeometric function (\ref{relation}); thus, the radial function (\ref{horizonsoluti}) reads
\begin{eqnarray}
\notag R(y) &=& C_1 e^{-(i\omega \ell \ln y)/\left(2 \sqrt{-k-12\nu \phi_0^2}\right)}(1-y)^\beta\frac{\Gamma(c)\Gamma(c-a-b)}{\Gamma(c-a)\Gamma(c-b)} {_2}F{_1}(a,b,a+b-c;1-y)  \\
  && + C_1 e^{-(i\omega \ell \ln y)/\left(2 \sqrt{-k-12\nu \phi_0^2}\right)}(1-y)^{3/2-\beta}\frac{\Gamma(c)\Gamma(a+b-c)}{\Gamma(a)\Gamma(b)} {_2}F{_1}(c-a,c-b,c-a-b+1;1-y)\,,
\end{eqnarray}
and at infinity, it can be written as
\begin{equation}\label{asintotica}\
R_{asymp.}(y) = C_1 (1-y)^\beta \frac{\Gamma(c)\Gamma(c-a-b)}{\Gamma(c-a)\Gamma(c-b)}+ C_1 (1-y)^{3/2-\beta}\frac{\Gamma(c)\Gamma(a+b-c)}{\Gamma(a)\Gamma(b)}~.
\end{equation}
Now, considering $\beta=1/2$ or $\beta=1$, we obtain that, similar to the previous case, both terms of this expression go to zero for $y \rightarrow 1$. So, the Dirichlet boundary condition is automatically satisfied for any value of $\omega$, which implies a continuous spectrum and instability. However, we will show that a discrete set of QNFs, depending on the physical parameters of the black hole and the test field, can be obtained by considering Neumann boundary condition.

{\it Neumann boundary condition.} It is also possible to consider that the flux of the scalar field vanishes at spatial infinity, which implies the Neumann boundary condition.
The flux (\ref{flujo}) at infinity is given by
\begin{equation}
F(y \rightarrow 1) \propto -\frac{ |C_1|^2 r_{+}^3 (3/2-2\beta)}{\ell^2} Im (A_1^{\ast}A_2)
\end{equation}
where we have used the Eq. (\ref{asintotica}) to obtain the above expression, and $A_1$ and $A_2$ are given by
\begin{equation}
A_1=\frac{\Gamma(c)\Gamma(c-a-b)}{\Gamma(c-a)\Gamma(c-b)}\, , \,\,\,\,\, A_2= \frac{\Gamma(c)\Gamma(a+b-c)}{\Gamma(a)\Gamma(b)} ~.
\end{equation}
The flux vanishes at infinity if $c-a=-n$, $c-b=-n$, $a=-n$ or $b=-n$ for $n=0,1,2,...$. Therefore, the discrete frequencies for the AdS black holes are given by
\begin{equation}
\omega_1= \pm \frac{\sqrt{k+4 \kappa^2 -4 \nu \phi_0^2}}{2 \ell}-i \frac{ \sqrt{-k-12 \nu \phi_0^2} (3+4n)}{2 \ell}~, \,\,\,\,\, \omega_2= \pm \frac{\sqrt{k+4 \kappa^2 -4 \nu \phi_0^2}}{2 \ell}-i \frac{\sqrt{-k-12 \nu \phi_0^2} (1+4n)}{2 \ell}~,
\end{equation}
for $\beta=1/2$ or $\beta=1$.
Both sets of QNFs can be written concisely as:
\begin{equation}
\omega= \pm \frac{\sqrt{k+4 \kappa^2 -4 \nu \phi_0^2}}{2 \ell}-i \frac{ \sqrt{-k-12 \nu \phi_0^2} (2n+1)}{2 \ell}~,
\end{equation}

The imaginary part of the QNFs is always negative for $k=1,0,-1$ and $\nu<0$; so, in these cases the propagation of a scalar field is formally stable in this background. However, for the topological black holes with $k=-1$ and $\nu > \xi^2 /\phi_0^2$ the QNFs are purely imaginary, and if the parameter $\xi^2$ takes values in the range $0 \leq \xi^2<4\nu \phi_0^2-1/4$, the imaginary part of the fundamental quasinormal frequency is positive, implying that the propagation of a scalar field on the topological black hole is unstable.

\subsection{Quasinormal modes for Lifshitz black holes with $z=4$}

It seems difficult to obtain analytical solutions to the radial equation (\ref{radial}) in the background described by (\ref{metric3}). However, it is worth mentioning that the radial equation at infinity $r \rightarrow \infty$ can be written as
\begin{equation}
r^2 R''(r)+7 r R'(r)+9 R(r)=0 ~,
\end{equation}
whose solution is $R_{asymp.} (r)  =C_1/r^3+C_2 \ln (r) /r^3$, and the field goes to zero at infinity. 
Therefore, the Dirichlet boundary condition is automatically satisfied for any value of $\omega$, and there is no discrete set of QNFs. This is similar to what occurs for the AdS black holes discussed before. Probably, a discrete set of QNFs exists when considering a different boundary condition, such as Neumann boundary condition, and the obtention of it could require of numerical techniques.

\section{Final Remarks}

We considered a gravitating system consisting of a conformally invariant scalar field coupled to four-dimensional Conformal Weyl Gravity. We found a new class of exact four-dimensional asymptotically AdS and Lifshitz black hole solutions with spherical, flat and hyperbolic event horizons for a certain range of values of the parameters. They are characterized by a scalar field with a power law behavior, which is regular everywhere outside the event horizon and vanishes at spatial infinity. The Lifshitz black hole solutions presented here are characterized by the dynamical exponents $z=0$ and $z=4$. Then, we discussed the thermodynamics of the solutions. The calculation of the mass was performed by using the generalized off-shell Noether potential method \cite{Kim:2013zha}, which has been applied to CG and also to other higher derivative gravity theories. In addition, we computed the Wald entropy of the solutions, and we showed that all the black holes satisfy the first law of black hole thermodynamics. It was found that the AdS black holes and the Lifshitz black holes with $z=0$ have zero mass and zero entropy, although they have a non-zero temperature; therefore, there is no conserved charge associated with the integration constants, which can be interpreted as a kind of gravitational hair. A similar behavior also occurs with black hole solutions (with planar horizon) obtained in other gravity theories containing higher curvature terms in the action, such as Lovelock gravity and three-dimensional New Massive Gravity with non-minimally coupled scalar fields. On the other hand, we found that the Lifshitz black hole with $z=4$ has non-zero values of the conserved quantities and thus more interesting thermodynamics. Also, we studied the horizon thermodynamics of the Lifshitz black hole solutions with $z=4$, and we demonstrated that the first law of black hole thermodynamics emerges from the field equations evaluated on the horizon. Furthermore, we studied the propagation of a conformally coupled scalar field in these backgrounds and we calculated the QNMs analytically in several cases. We found that for AdS black holes and Lifshitz black holes with $z=4$, the Dirichlet boundary condition is automatically satisfied for any value of $\omega$, implying a continuous spectrum. This kind of behavior has been observed in other AdS and Lifshitz black hole geometries for scalar type electromagnetic field perturbations \cite{Lopez-Ortega:2014oha, Gomez-Navarro:2017fyx}, see also Refs. \cite{ Estrada-Jimenez:2013lra, Cruz:2015nza} where continuous spectrum of frequencies have been obtained for a two-dimensional black hole. Nevertheless, it is expected that the QNFs depend on the physical parameters that describe the black hole and the test field. In \cite{Lopez-Ortega:2014oha, Gomez-Navarro:2017fyx} a different boundary condition was proposed, by canceling the leading term of the asymptotic behavior of the radial function well defined QNFs were obtained. We showed that discrete sets of well defined QNFs, which depend on the physical parameters of the black hole and the test field, can be obtained by considering that the flux of the scalar field vanishes at spatial infinity (Neumann boundary conditions). We obtained analytical expressions for the QNFs of AdS black holes and Lifshitz black holes with $z=0$. We found that the discrete QNFs can have real and imaginary parts, and the imaginary part is always negative for black holes with spherical and flat horizon $k=1,0$; so, in these cases the propagation of a scalar field is formally stable in these backgrounds. However, we found that for a topological AdS black hole (with $k+12\nu \phi_0^2=0$), and for the topological $z=0$ Lifshitz black hole, if the parameter $\xi^2$ takes values in the range $0 \leq \xi^2<4\nu \phi_0^2-1/4$, the imaginary part of the fundamental quasinormal frequency is positive, implying that the propagation of a scalar field on the topological black hole is unstable. Analogously, for a topological AdS black hole, with $A=0$, the propagation of a scalar field is unstable if $0 \leq \xi^2 < 1/12-2 (r_+ / \ell)^2$. It would be interesting to explore the possible holographic applications to condensed matter systems of the solutions found here and also to extend this study to a gravitating system with a charged scalar field.

\section*{Acknowledgments}
 
We would like to thank the anonymous referee for valuable comments which help us to improve the quality of our paper. F.H. would like to thank P.A. Gonz\'alez. Y.V. would like to thank G. Giribet for useful comments. This work was partially funded by the Comisi\'on Nacional de Ciencias y Tecnolog\'ia through FONDECYT Grant 11140674 (F.H.) and by the Direcci\'{o}n de Investigaci\'{o}n y Desarrollo de la Universidad de La Serena (F.H. and Y.V.).

\appendix

\end{document}